\documentclass[a4paper]{jpconf}
\usepackage{epsfig}
\usepackage{units}     
\usepackage{mathrsfs}   
\usepackage{graphicx}    
\usepackage{dcolumn}    
\usepackage{bm}             
\usepackage{amsmath,amsfonts,amssymb}
\usepackage{graphicx,epic,eepic}
\usepackage{subfigure}
\usepackage{graphics}
\usepackage[dvips]{color}
\newcommand{\refb}[1]{(\ref{#1})}

\def\diag{\mathrm{diag}}
\def\tx{(t,\mathbf{x})}
\def\txp{(t,\mathbf{x}')}

\newcommand{\comBig}[2]{\Big[ #1,#2 \Big]}

\newcommand{\op}[1]{\hat{#1}}

\def\eg{{\emph{e.g.}}}
\def\d{{\mathrm{d}}}

\begin{document}
%%%%%%%%%%%%%%%%%%%%%%%%%%%%%%%%%%%%%%%%%%%%%%%%%%%
%
\title{Signature-change events in emergent spacetimes with anisotropic scaling}
%
%%%%%%%%%%%%%%%%%%%%%%%%%%%%%%%%%%%%%%%%%%%%%%%%%%%
%
\author{Silke Weinfurtner$^1$, Angela White$^2$, and
Matt Visser$^3$}
\address{$^1$  Department of Physics and Astronomy, University of British Columbia, \\
\ \ Vancouver, BC, V6T 1Z1, Canada}
\address{$^2$ School of Mathematics and Statistics, Newcastle University\\
\  \ Herschel Building, Newcastle upon Tyne, NE1 7RU, United Kingdom}
\address{$^3$ School of Mathematics, Statistics, and Operations Research, \\
\ \ Victoria University of Wellington, PO Box 600, Wellington, New Zealand}
\ead{silke@physics.ubc.ca,  angela.white@ncl.ac.uk, matt.visser@msor.vuw.ac.nz}
\date{\TODAY; \LaTeX-ed \DayOfWeek, \today; \daytime}   
%
%
%%%%%%%%%%%%%%%%%%%%%%%%%%%%%%%%%%%%%%%%%%%%%%%%%%%
%
\begin{abstract}
We investigate the behaviour of quantum fields coupled to a spacetime geometry exhibiting finite regions of Euclidean (Riemannian) signature. Although from a gravity perspective this situation might seem somewhat far fetched, we will demonstrate its direct physical relevance for an explicitly realizable condensed matter system whose linearized  perturbations experience an effective emergent spacetime geometry with externally controllable signature. This effective geometry is intrinsically quantum in origin, and its signature is determined by the details of the microscopic structure. At the level of the effective field theory arising from our condensed matter system we encounter explicit anisotropic scaling in time and space. Here Lorentz symmetry is an emergent symmetry in the infrared. This anisotropic scaling of time and space cures some of the technical problems that arise when working within a canonical quantisation scheme obeying strict Lorentz invariance at all scales, and so is helpful in permitting signature change events to take place. 
\end{abstract}
%
%%%%%%%%%%%%%%%%%%%%%%%%%%%%%%%%%%%%%%%%%%%%%%%%%%%
%
\section{Motivation\label{Sec:Motivation}}
%
%%%%%%%%%%%%%%%%%%%%%%%%%%%%%%%%%%%%%%%%%%%%%%%%%%%
Recently much attention has been paid to the possibility of ad-hoc construction of quantum field theories with a certain degree $z$ of anisotropic scaling between the space and time dimensions --- this anisotropic scaling often significantly improving the ultraviolet behaviour.  See for example~\cite{Horava:2008jf, Horava:2008ih, Horava:2009rt, Horava:2009uw} and~\cite{Visser:2009fg, Sotiriou:2009kx, Sotiriou:2009vn, Volovik:2009av, Charmousis:2009mz, Lu:2009em}.  Wide-scale interest in these ideas was largely ignited by Ho\v{r}ava's specific model for quantum gravity in $3+1$ dimensions, based on anistropic scaling at the $z=3$ Lifshitz point~\cite{Horava:2009rt, Horava:2009uw},  but the basic ideas have much wider applicability --- for instance it is possible to extend these ideas to general spatial dimensions~\cite{Horava:2008jf, Horava:2008ih,  Horava:2009rt, Horava:2009uw, Visser:2009fg, Sotiriou:2009kx, Sotiriou:2009vn}, such that the engineering dimensions of space and time are of degree $z=d$, such that
\begin{equation}
[\d x] = [\kappa]^{-1} \quad \mbox{and} \quad [\d t]=[\kappa]^{-z}.
\end{equation}
Here $d$ is the number of spatial dimensions, and $\kappa$ is a placeholder with dimensions of momentum. Based on the anisotropy between space and time it is possible to modify the Einstein--Hilbert action of general relativity, adding one kinetic and several higher order spatial curvature terms, containing up to $2d$ spatial derivatives. See for example~\cite{Horava:2009rt, Horava:2009uw, Sotiriou:2009kx,Sotiriou:2009vn}. It is worth emphasising that these models do not need (or even permit) more than $2$ time derivatives. There are strong indications that such a quantum gravity model based on anisotropic scaling exhibits vastly improved behaviour at small scales, which alters the theory to be renormalisable (possibly even finite), at the cost of giving up Lorentz symmetry as an exact symmetry~\cite{Horava:2009rt, Horava:2009uw}. 

Shortly after Ho\v{r}ava's papers on this quantum gravity candidate~\cite{Horava:2009rt, Horava:2009uw},  anisotropic scaling was adapted for use as a regulator for scalar polynomial quantum field theories, thereby throwing light on their renormalizability properties~\cite{Visser:2009fg}. The free Lagrangian (for a scalar field with mass $m_s$) in this proposal is given by
\begin{equation}
\label{Eq:Visser.Sfree}
S_\mathrm{free} = \int \left\{ \dot{\phi}^2 - \phi \left[ m_s^2 - c^2 \triangle + ... + (-\triangle)^z \right] \phi \right\} \, \d t \, \d^d x.
\end{equation}
It has been shown that in $d+1$ spacetime dimensions subject to $z=d$ anisotropic scaling arbitrary polynomial interactions of the form $P(\phi)=\sum_{n=1}^N g_n \,\phi^n$ are perturbatively ultraviolet finite when normal ordering is applied. Even without normal ordering this quantum field theory is perturbatively renormalizable.

As already mentioned above, in both proposals the ultraviolet modifications are more or less chosen ad-hoc,  with hindsight showing that one gets a better ultraviolet control over the quantum behaviour of the gravitational\,/\,scalar field. In this spirit we think it may be interesting to note that there is a physically realistic class of condensed matter systems that naturally exhibit some of the features predicted in equation~(\ref{Eq:Visser.Sfree}). A specific example of such a system is an ultra-cold gas of bosons in its superfluid phase, referred to as a Bose--Einstein condensate (BEC), or more generally any condensed matter system exhibiting a Bogoliubov dispersion relation. It has been well established, see~\cite{Unruh:1981bi, Barcelo:2005ln, Volovik:2003jn, Matt.-Visser:2002ot, Schutzhold:2007aa}, that small perturbations around the condensate experience an effective\,/\,acoustic\,/\,emergent spacetime geometry. In this context Lorentz symmetry is not an exact symmetry, it is emergent~\cite{Weinfurtner:2007qy, Weinfurtner:2007aa, Weinfurtner:2006iv, Liberati:2006kw, Weinfurtner:2006nl, Liberati:2006sj}. Previously, see~\cite{ Weinfurtner:2007aa, White:2008xr}, we have shown that the signature of the emergent spacetime geometry is related to the nature of the interaction between the fundamental bosons, and that it is possible to drive (sudden) finite transitions between periods of different spacetime signature. With certain technical assumptions (including a translation invariant background condensate) we can show that the linearised BEC resembles a scalar field with anisotropic scaling, as suggested in~\cite{Visser:2009fg},
\begin{equation}
S_\mathrm{BEC}^{(1)} \approx \int \left\{ \dot{\phi}^2 - \phi \left[  - c_0^2 \triangle + \gamma^2_{qp} \,  \triangle^2 \right] \phi \right\} \, \d t \, \d^d x.
\end{equation}
Here $c_0$ is the speed of sound (which takes the role of the speed of light $c$ within the context of emergent spacetimes), $\gamma_{qp}$ is associated with the microscopic substructure, the elementary bosons, while the low-momentum speed of sound $c_0$ depends on the condensate. However in a more general emergent spacetime background (especially once the background is not translation invariant) the effective free field action is considerably more complicated and has to be treated with care~\cite{ Weinfurtner:2007aa,  Weinfurtner:2007ab}. Since the degree of anisotropy is in this case (only) $z=2$, the modifications are not sufficient to cure all problems associated with finite regions of Euclidean (Riemannian) signature.
Without anisotropic scaling the particle number-density produced from finite regions of Euclidean (Riemannian) signature is given by
\begin{equation}
N_k = \sinh^2\left[  \sqrt{ m_s^2 +  k^2 } \; \Delta \tau \right].
\end{equation}
This corresponds to a particle spectrum that is exponentially increasing both in wave-number $k$, and in the length (``duration'') of the Euclidian region $\Delta \tau$. For a detailed discussion on the subject consult~\cite{ Weinfurtner:2007aa, White:2008xr}. This result shows that regardless of the number of spatial dimensions and the mass $m_s$ of the particle, the number-density and energy density of the excitations is infinite,  and thus the quantisation scheme at hand breaks down. We will see that adding anisotropic scaling in the ultraviolet greatly improves the situation.

%%%%%%%%%%%%%%%%%%%%%%%%%%%%%%%%%%%%%%%%%%%%%%%%%%%
%
\section{Emergent spacetime with anisotropic scaling\label{Sec:EmergentAnisotropic}}
%
%%%%%%%%%%%%%%%%%%%%%%%%%%%%%%%%%%%%%%%%%%%%%%%%%%%

In the following discussion we will introduce a ``real life'' example where the notion of geometry is emergent. The emergent spacetime we are going to study closely involves: (i) a microscopic system consisting of an ultra-cold weakly interacting gas of bosons; (ii) which exhibits (after a first-order phase transition) a mean-field regime where the microscopic degrees of freedom give way to collective variables; (iii) small perturbations above the ground state which are dominated by a `geometrical object', a symmetric second rank tensor; (iv) Lorentz symmetry as an emergent symmetry in the infrared, while at higher energies the effective field theory shows anisotropic scaling in time and space; (v) the signature of the emergent spacetime is of microscopic origin and can be `switched' by the application of external magnetic fields.

%+++++++++++++++++++++++++++++++++++++++++++++++++++++++++++++++++
\subsection*{\textsl{(i) Microscopic system}\label{Sec:}}
%+++++++++++++++++++++++++++++++++++++++++++++++++++++++++++++++++
The microscopic degrees of freedom at hand are represented by individual quantum field operators that act on quantum states (Hilbert space of states; \eg, Fock space), creating, $\op{\psi}^{\dag}\tx$, or destroying, $\op{\psi}\tx$, an individual boson at a particular point in space and time. These operators satisfy the usual boson commutators~\cite{Pethick:2001aa,Gardiner:2000aa}:
\begin{eqnarray} \label{Eq:com}
\comBig{\op{\psi}\tx }{\op{\psi}\txp}=\comBig{\op{\psi}^{\dag}\tx}{\op{\psi}^{\dag}\txp}=0\,; \quad
\comBig{\op{\psi}\tx}{\op{\psi}^{\dag}\txp}=\delta(\mathbf{x}-\mathbf{x}')\,.
\end{eqnarray}
The Hamiltonian of the microscopic system is chosen to be non-relativistic,
\begin{eqnarray} \label{Eq:Hamiltonian}
\op{H}= \int{ d\mathrm{x} \left(
- \op{\psi}^{\dag} \frac{\hbar^{2}}{2m}\nabla^{2} \op{\psi}
+ \op{\psi}^{\dag} V_{\mathrm{ext}} \op{\psi}
+ \frac{U}{2} \, \op{\psi}^{\dag}\op{\psi}^{\dag}\op{\psi}\op{\psi} \right) } \, ,
\end{eqnarray} 
and obeys $SO(2)$ symmetry transformations of the form $ \op{\psi}  \to  \op{\psi} \, \exp(i \alpha)$.
The Hamiltonian is a sum of  the kinetic energy of the boson field, and 
two potential energy contributions: the external trap potential 
$V_{\mathrm{ext}}$, and the quasi-particle interactions. The extreme
dilution of the gas (typically $10^{13}-10^{15}\,
\mathrm{atoms}/\mathrm{cm}^{3}$) suppresses interactions involving more than two particles, and in the weakly interacting regime the actual
inter-atom potential has been approximated by a pseudo-contact
potential
\begin{equation} \label{Eq:U}
U = \frac{4 \pi \hbar^{2} a}{m}\,.
\end{equation}
Here $m$ is the single-boson mass, and $a$ the $s$-wave scattering length.

%+++++++++++++++++++++++++++++++++++++++++++++++++++++++++++++++++
\subsection*{\textsl{(ii) Mean-field regime}\label{Sec:}}
%+++++++++++++++++++++++++++++++++++++++++++++++++++++++++++++++++
Under an appropriate cooling process (one that protects the gas from solidification through a first-order phase-transition) a new state of matter will occur, the Bose--Einstein condensate. The condensate is a complex-valued macroscopic mean-field that is a result of a spontaneous breaking of the $SO(2)$ symmetry, such that the field operators for topologically trivial regions (without zeros or singularities) acquire non-zero expectation values,
\begin{equation} \label{Eq:psi}
\langle \op{\psi}\tx \rangle =
\psi\tx = \sqrt{n_{0}\tx} \; \exp\left(i \, \phi_{0}\tx\right) \neq 0 .
\end{equation}
The individual microscopic degrees of freedom give way to collective variables, such as the condensate density $n_0 \equiv  n_0\tx$, and the phase $\phi_0 \equiv \phi_0 \tx$~\cite{Pethick:2001aa, Gardiner:2000aa}.

%+++++++++++++++++++++++++++++++++++++++++++++++++++++++++++++++++
\subsection*{\textsl{(iii) Emergent metric tensor}\label{Sec:}}
%+++++++++++++++++++++++++++++++++++++++++++++++++++++++++++++++++
%
The emergence of a geometrical rank two tensor enters (in a by now quite standard manner) at the linear level after an expansion around the mean-field variables~\cite{Barcelo:2002dp}. In terms of the hermitian density $\op{n}$ and phase $\op{\phi}$ fluctuations we have~\cite{Weinfurtner:2007qy,Weinfurtner:2007ab,Weinfurtner:2007aa}:
\begin{eqnarray}
\label{Eq:exp.psi_1}
 \psi\tx + \delta \op{\psi}\tx &\to&   \psi\tx \, \left( 1+ \frac{1}{2}\frac{\op{n}}{n_{0}} + i \op{\phi}  \right)\,, \\
\label{Eq:exp.psi_2}
\psi^*\tx + \delta\op{\psi}^{\dag}\tx & \to & \psi^{*}\tx \, \left(1 + \frac{1}{2}\frac{\op{n}}{n_{0}} - i \op{\phi} \right)\,.
\end{eqnarray}
Form the usual boson commutators of equation~(\ref{Eq:com}), it is straightforward to show that the new operators are a pair of canonical variables:
\begin{eqnarray}
\comBig{\op{n}\tx }{\op{n}\txp}=\comBig{\op{\phi}\tx}{\op{\phi}\txp}=0\,; \quad \mbox{and} \quad
\comBig{\op{n}\tx}{\op{\phi}\txp}=i\delta(\mathbf{x}-\mathbf{x}')\,.
\label{Eq:n_psi_com2}
\end{eqnarray}
By assuming the perturbation to be
small, allowing us to neglect quadratic and higher-order products
of the perturbation field $\delta \hat \psi$, and working within the hydrodynamic limit (`long-wavelength' limit), for low-momentum modes we are able to express the excitation spectrum in a very compact and insightful way:
\begin{equation} \label{Eq:KLG.1}
\frac{1}{\sqrt{\left\vert \det(g_{ab}) \right\vert}} \; \partial_{a} \left( \sqrt{ \left\vert \det(g_{ab}) \right\vert }\; g^{ab} \; \partial_{b}   \hat \phi \right) =0 \, . \\
\end{equation}
Here we have  introduced the ``acoustic metric''
\begin{equation} \label{Eq:g.1}
g_{ab} = \left( \frac{c_{0}}{U/\hbar} \right)^{\frac{2}{d-1}}
\left[
\begin{array}{cccc}
-\left(c_{0}^{2}-\mathbf{v}^{2}\right) & -v_{x} & -v_{y} & -v_{z} \\
-v_{x} & 1 & 0 & 0 \\
-v_{y} & 0 & 1 & 0 \\
-v_{z} & 0 & 0 & 1
\end{array}
\right]\, ,
\end{equation}
a covariant rank two metric tensor, whose entries are purely collective variables. For a detailed derivation see~\cite{Weinfurtner:2007qy, Weinfurtner:2007aa, Weinfurtner:2007ab}. The conformal factor depends on the spatial dimensionality, $d$, of the condensate cloud. Here we have introduced the background velocity of the condensate as
\begin{equation}
\mathbf{v} = \frac{\hbar}{m} \nabla \phi_{0},
\end{equation}
and the propagation speed for excitations within this hydrodynamic approximation simplifies to a common sound speed given by
\begin{equation} \label{Eq:sound.speed.hydro}
c_{0}^{2} = \frac{n_{0}U}{m}\,.
\end{equation}
In the same manner that we eliminated all (explicit) occurrences of density
perturbations in the excitation spectrum, it is possible to show that the commutation relation~\refb{Eq:n_psi_com2} can be re-written solely in terms of the field $\hat \phi$ and its conjugate momentum $\hat \Pi_{\hat \phi}$ on the effective curved spacetime geometry:
\begin{equation} \label{Eq:Pi_psi_com2}
\comBig{\op{n}\tx}{\op{\phi}\txp}=i\delta(\mathbf{x}-\mathbf{x}') \to \comBig{\hat\phi\tx}{\hat \Pi_{\hat \phi}\txp}=i \delta(\mathbf{x}-\mathbf{x}')\,.
\end{equation}

%+++++++++++++++++++++++++++++++++++++++++++++++++++++++++++++++++
\subsection*{\textsl{(iv) Emergent Lorentz symmetry and anisotropic scaling}\label{Sec:}}
%+++++++++++++++++++++++++++++++++++++++++++++++++++++++++++++++++
Within the context of emergent spacetimes Lorentz symmetry is not an exact symmetry. Low-energy\,/\,infrared excitations around the macroscopic field, which describes the vacuum state of the system, experience an emergent Lorentz symmetry. This symmetry will be broken in the high-energy\,/\,ultraviolet regime, that is at scales dominated by the underlying microscopic theory. These corrections are of a non-perturbative nature involving only higher spatial derivatives~\cite{Weinfurtner:2007qy, Weinfurtner:2007aa, Weinfurtner:2006iv, Liberati:2006kw, Weinfurtner:2006nl, Liberati:2006sj,  Weinfurtner:2007ab}.

It is possible to incorporate these non-perturbative corrections into the derivation of the effective spacetime geometry by formally replacing the interaction strength $U$ with an appropriate  differential operator:
\begin{equation}
\label{Eq.Effective.U}
U \to \widetilde U =  U - \frac{\hbar^2}{2 m}\;\widetilde D_{2}\, ;    \quad \mbox{where} \quad
\widetilde D_{2} =\frac{1}{2} \left\{ \frac{(\nabla n_{0})^{2} -(\nabla^{2}n_{0})n_{0}}{n_{0}^{3}} -\frac{\nabla n_{0}}{n_{0}^{2}}\nabla +\frac{1}{n_{0}}\nabla^{2} \right\}.
\end{equation}
This replacement now leads to an integro-differential equation,
\begin{equation}
\label{Eq:WaveEquation}
\partial_{a} \left(f^{ab}\; \partial_{b} \hat \phi \right) = 0 \, ,
\end{equation}
where we have introduced a matrix $f^{ab}$ with inverse-differential-operator-valued entries:
\begin{equation}
\label{Eq:f}
f^{ab} = \hbar
\left[ \begin{array}{c|c} \vphantom{\Big|} -\widetilde U^{-1} & -\widetilde U^{-1} v^{j} \\ \hline \vphantom{\Big|} -v^{i} \widetilde U^{-1} & \frac{n_{0}}{m}\delta^{ij}- v^i \widetilde U^{-1} v^j \end{array}\right] \, .
\end{equation}
In general $\widetilde U^{-1}$ is an integral operator which can be formally expanded as~\footnote{The \emph{formal} series converges only on the subspace of functions spanned by the eigenfunctions whose eigenvalues satisfy
\begin{equation}
\nonumber
\lambda \left( \frac{\hbar^{2}}{2m} U^{-1} \widetilde D_{2} \right) < 1 \, .
\end{equation}
} 
\begin{eqnarray}
\label{Eq:Integral.Differential.Operator}
\widetilde U^{-1} = U^{-1} &+& \left[\frac{\hbar^2}{2m}\right] \, U^{-1} \widetilde D_{2} \, U^{-1}  \\ \nonumber
 &+& \left[\frac{\hbar^2}{2m}\right]^{2} \, U^{-1} \widetilde D_{2} \,  U^{-1} \widetilde D_{2} \, U^{-1} \\ \nonumber
&+& \left[\frac{\hbar^2}{2m}\right]^{3} \, U^{-1} \widetilde D_{2} \, U^{-1} \widetilde D_{2} \,  U^{-1} \widetilde D_{2} \, U^{-1} + ...
 \end{eqnarray}
Note that we additionally require that there
exists an (inverse) metric tensor $g^{ab}$ such that
\begin{equation}
\label{Eq:EmergentGeometry}
f^{ab} \equiv \sqrt{-g} \, g^{ab} \, ,
\end{equation}
where $g$ is the determinant of $g_{ab}$. Then the
connection is formally made to the field equation for a minimally
coupled massless scalar field in a curved spacetime; see equation~\refb{Eq:g.1}.
Let us rewrite this  (for more details see~\cite{Weinfurtner:2007ab}) as
\begin{equation}
\widetilde U \to  U \left(1- \frac{\gamma_{qp}^2}{c_0^2}\;\widetilde D \right)\, ;    \quad \mbox{where} \quad
\widetilde D = \left\{ \left(\frac{\nabla n_{0}}{n_{0}}\right)^2 - \frac{\nabla^{2}n_{0}}{n_{0}} -\frac{\nabla n_{0}}{n_{0}}\nabla + \nabla^{2} \right\},
\end{equation}
and further take $n_0 \tx \equiv \mbox{const}$ and $U\tx \equiv U(t)$, so that the background is translation invariant. We then get $c_0\tx \equiv c_0(t)$, and see that
\begin{equation}
\widetilde U \to  U \left(1- \frac{\gamma_{qp}^2}{c_0^2}\; \nabla^{2} \right)\, ,
\end{equation}
and
\begin{equation}
\widetilde U^{-1} \to  U^{-1}  \left( 1 - \sum_{n=1}^{\infty} \left[ \frac{\gamma_{qp}}{c_0}  \right]^{2n} \, \triangle^n  \right).
\end{equation}
Here $\gamma_{qp}$ is constant and only depends on the mass of the fundamental bosons:
\begin{equation}
\gamma_{qp} = \frac{\hbar}{2m}.
\end{equation}
Therefore it can be seen that the whole notion of an emergent spacetime geometry is, even under this very simplified situation, difficult to set-up if we are outside the validity of the Lorentz invariant regime, and one practically has to give up the notion of geometry. However, in cases of an isotropic spacetime, such as FRLW spacetime it is very simple to construct the effective field theory, since a mapping onto momentum space is trivial, $\triangle \to - k^2$, and
\begin{equation}
\widetilde U^{-1} \to  U^{-1}  \left( 1 - \sum_{n=1}^{\infty} \left[ \frac{\gamma_{qp}}{c_0}  \right]^{2n} \, (-k^2)^n  \right) \to \frac{1}{U \left(1+ \frac{\gamma_{qp}^2}{c_0^2} k^2 \right)}.
\end{equation}
Altogether for $v^i  \equiv 0$ on a translation invariant background, we have
\begin{equation}
f^{ab} = \hbar \,  \diag \left( - \widetilde U^{-1} , \frac{n_0}{m}, \frac{n_0}{m},\frac{n_0}{m} \right).
\end{equation}
For the equation of motion we see
\begin{equation}
\partial_t \partial_t \hat \phi + (f^{tt})^{-1} \, f^{ij} \partial_i \partial_j \hat \phi = 0,
\end{equation}
which we re-write as
\begin{equation}
\partial_t \partial_t \hat \phi - \left( c_0^2 \, \triangle  - \gamma_{qp}^2 \, \triangle^2 \right) \hat \phi = 0.
\end{equation}
Subject to these  approximations,  and the simplifications imposed on the condensate background,  we are able to associate the effective field theory of our BEC system with the specific case of anisotropic scaling at $z=2$:
\begin{equation}
S_\mathrm{BEC}^{(1)} \approx \int \left\{ \dot{\phi}^2 - \phi \left[  - c_0^2 \triangle + \gamma^2_{qp} \,  \triangle^2 \right] \phi \right\} \, \d t \, \d^d x.
\end{equation}
A change to momentum space leads to the well-known Bogoliubov dispersion relation
\begin{equation}
\omega^2 = c_0^2 \, k^2 + \gamma^{2}_{qp} \, k^4.
\end{equation}
Therefore we are able to connect the anisotropic scaling at $z=2$ with an emergent Lorentz symmetry for momenta below the effective Lorentz breaking scale,
\begin{equation}
k^2 \ll \frac{c_0^2}{\gamma_{qp}^2}.
\end{equation}
For further information about the role of Lorentz symmetry breaking within the emergent spacetime\,/\,analogue models for gravity\,/\,acoustic metric programme see~\cite{Weinfurtner:2007qy,Weinfurtner:2006iv,Liberati:2006kw,Weinfurtner:2006nl,Liberati:2006sj, Visser:1993tk, Visser:1997ux}.

%+++++++++++++++++++++++++++++++++++++++++++++++++++++++++++++++++
\subsection*{\textsl{(v) Signature and signature change events}\label{Sec:}}
%+++++++++++++++++++++++++++++++++++++++++++++++++++++++++++++++++
The signature of the emergent spacetime geometry can easily be read off by setting $v^i \to 0$. This can always be done locally, and we can read off the pattern of metric eigenvalues from equation~(\ref{Eq:g.1}):
\begin{equation} \label{Eq:eta.1}
\eta_{ab} \sim
\left[
\begin{array}{cccc}
-c_{0}^{2} & 0 & 0 & 0 \\
0 & 1 & 0 & 0 \\
0 & 0 & 1 & 0 \\
0 & 0 & 0 & 1
\end{array}
\right]\, .
\end{equation}
Since $c_0^2 \sim U$ we are able to relate the signature of the emergent spacetime geometry with the qualitative behaviour of the atom-atom interactions, here approximated with a purely contact potential only accounting for $s$-wave scattering processes~\cite{Weinfurtner:2007qy, Weinfurtner:2007aa, Weinfurtner:2007ab}. (Certainly this particular approximation is a significant weakness of the whole ansatz,  and it would be interesting to extend the analysis towards a more realistic interaction potential.) 
Experimentally it is possible to drive transitions between negative and positive values of $U$, where
\begin{equation} \label{Eq:nature_of_a}
\begin{array}{rl}
U > 0 & \quad\mbox{repulsive}\, ; \\[5pt]
U < 0 & \quad \mbox{attractive}\, .
\end{array}
\end{equation}
Such changes in the \emph{sign} of the interaction are accessible by tuning external magnetic fields that interact with the inter-atomic potential; this process is called Fesh\-bach resonance \cite{Inouye:1998aa}. This is not just an  ``in principle'' possibility, since in 2001 a BEC experiment~\cite{Donley:2001aa,Roberts:2001aa} was carried out that can be viewed as the first emergent spacetime experiment involving particle production resulting from a signature change event, see~\cite{Weinfurtner:2007aa, White:2008xr}.
 For a detailed discussion regarding the behaviour of the effective field theory exposed to a finite region of Riemannian signature consult~\cite{Weinfurtner:2007aa, White:2008xr}. There it has been shown that at very high momenta, due to the anisotropic scaling of time and space, the number of quasi-particles being produced in sudden signature variations of the form Lorentzian--Euclidean--Lorentzian, $(-,+++) \to (+,+++) \to  (-,+++)$,  scales asymptotically as $N_{k} \sim k^{-4}$.  Therefore
\begin{equation}
N_{\leqslant k}\sim 2^{d-1}\pi \, \int dk \,k^{d-1} \, k^{-4} \sim k^{d-4} .
\end{equation}
That is, the total number of quasi-particles produced is finite, as expected on physical grounds. However, it is easy to see that the total energy emitted,
\begin{equation} \label{Eq.total.engery.general}
E_{\leqslant k}=2^{d-1}\pi \, \int_{k} dk \,k^{d-1} \; N_{k} \, \omega_{k}  \sim  2^{d-1}\pi \, \int_{k} dk \, k^{d-1} \, k^{-2} \sim k^{d-2},
\end{equation}
is still infinite.  This interesting effect is due to the specific behaviour of the coefficients characterizing  the anisotropic scaling --- only the $k^2$-term in the effective Lagrangian is affected by the Feschbach resonance, while the $k^4$-term remains --- at the linear level --- unaffected:
\begin{equation}
 \left\{ \dot{\phi}^2 - \phi \left[  - c_0^2 \triangle + \gamma^2_{qp} \,  \triangle^2 \right] \phi \right\} 
\to
 \left\{ \dot{\phi}^2 - \phi \left[  + c_0^2 \triangle + \gamma^2_{qp} \,  \triangle^2 \right] \phi \right\} 
\to 
 \left\{ \dot{\phi}^2 - \phi \left[  - c_0^2 \triangle + \gamma^2_{qp} \,  \triangle^2 \right] \phi \right\} .
\end{equation}
Therefore at energy scales $k^2  \gg c_0^2/\gamma_{qp}^2$ the higher spatial derivatives (which in fact define the $z=2$ Lifschitz point) are unaffected:
\begin{equation}
 \left\{ \dot{\phi}^2 - \phi \left[   \gamma^2_{qp} \,  \triangle^2 \right] \phi \right\} 
\to
 \left\{ \dot{\phi}^2 - \phi \left[   \gamma^2_{qp} \,  \triangle^2 \right] \phi \right\} 
\to 
 \left\{ \dot{\phi}^2 - \phi \left[   \gamma^2_{qp} \,  \triangle^2 \right] \phi \right\} .
\end{equation}
However we would like to emphasise that this linear regime is only applicable to some limited extent, since ultimately  three-body losses (and consequently backreaction effects) will play a rapidly growing role with the increasing duration of the attractive interaction\,/\,Riemannian signature regime.

%%%%%%%%%%%%%%%%%%%%%%%%%%%%%%%%%%%%%%%%%%%%%%%%%%%
%
\section{Summary and conclusions\label{Sec:Summary}}
%
%%%%%%%%%%%%%%%%%%%%%%%%%%%%%%%%%%%%%%%%%%%%%%%%%%%
We have discussed the concept of a spacetime emergent from a Bose--Einstein condensate, and the possibility of signature change events within this context~\cite{Weinfurtner:2007aa,  White:2008xr, Calzetta:2005yk, Calzetta1:2003xb, Hu:2003}. The recent interest in quantum gravity models sacrificing Lorentz symmetry as an exact symmetry  in order to obtain better control of the ultraviolet behaviour of quantum gravitational\,/\,field excitations ~\cite{Horava:2008jf,   Horava:2008ih, Horava:2009rt, Horava:2009uw, Visser:2009fg, Sotiriou:2009kx, Sotiriou:2009vn,Volovik:2009av,Charmousis:2009mz,Lu:2009em}, has motivated us to emphasise some of the features that `anisotropic scaling' and the presence of Lorentz-violating Lifschitz points may contribute to the subject of condensed-matter emergent spacetimes. We have, in brief, demonstrated how spacetime geometry --- as seen by quasiparticle excitations --- emerges from a mean-field structure that is truly quantum in origin~\cite{Barcelo:2005ln}. Although this model has no direct connection or application to the search of a quantum theory of gravity, we have shown that it nevertheless exhibits some (but not all) of the features that are essential to Ho\v{r}ava--Lifschitz gravity. The key features are that Lorentz symmetry is an emergent symmetry in the infrared, and that the non-perturbative terms added at the level of the effective field theory are anisotropic in space and time, involving only higher spatial derivatives. In fact, these two features seem to be common to all known emergent spacetimes, independent of the specific media from which they are emergent --- \emph{e.g.}, water, superfluids, electromagnetic waveguides, \emph{etc}.

Coming back to the specific model at hand, the Bose--Einstein condensate, we would like to point out that the free-field description for small perturbations is only an approximation. We are, strictly speaking, dealing with an interacting field theory, but due to the anisotropic scaling naturally built into the effective field theory the ultraviolet regime has some advantages compared to conventional field theories. A specific way to explore this is by exposing the effective field theory to one of the most drastic gravitational scenarios possible (besides curvature singularities): signature change events. While for a $z=1$ (Lorentz invariant) scaling of time and space the canonical quantisation scheme applied to quantum field theories on Riemannian manifolds breaks down, predicting infinite quasi-particle production per unit volume, we see that in contrast  $z=2$ anisotropic scaling is already sufficient to render quasiparticle production finite. It is straightforward to show that at $z=3$ the energy density per volume will also be finite~\cite{Weinfurtner:2007qy, Weinfurtner:2007aa, Weinfurtner:2006iv, Weinfurtner:2007ab,Jain:2006ki}.

The main lesson to be taken from the discussion is that nature quite naturally supplies us with little role models equipped with both of the desired features, both emergent Lorentz symmetry and anisotropic scaling.

%%%%%%%%%%%%%%%%%%%%%%%%%%%%%%%%%%%%%%%%%%%%%%%%%%%
%
\section*{References}
%
%%%%%%%%%%%%%%%%%%%%%%%%%%%%%%%%%%%%%%%%%%%%%%%%%%%

%
%%%%%%%%%%%%%%%%%%%%%%%%%%%%%%%%%%%%%%%%%%%%%%%%%%%
%%%%%%%%%%%%%%%%%%%%%%%%%%%%%%%%%%%%%%%%%%%%%%%%%%%
%%%%%%%%%%%%%%%%%%%%%%%%%%%%%%%%%%%%%%%%%%%%%%%%%%%

\begin{thebibliography}{10}

\bibitem{Horava:2008jf}
Petr Ho\v{r}ava.
\newblock {Quantum Criticality and Yang-Mills Gauge Theory}.
\newblock arXiv:0811.2217, 2008.

\bibitem{Horava:2008ih}
Petr Ho\v{r}ava.
\newblock {Membranes at Quantum Criticality}.
\newblock {\em JHEP}, 03:020, 2009.

\bibitem{Horava:2009rt}
Petr Ho\v{r}ava.
\newblock Spectral dimension of the universe in quantum gravity at a {Lifshitz}
  point.
\newblock {\em Physical Review Letters}, 102(16):161301, 2009.

\bibitem{Horava:2009uw}
Petr Ho\v{r}ava.
\newblock {Quantum Gravity at a Lifshitz Point}.
\newblock {\em Phys. Rev.}, D79:084008, 2009.

\bibitem{Visser:2009fg}
Matt Visser.
\newblock {Lorentz symmetry breaking as a quantum field theory regulator}.
\newblock arXiv:0902.0590, 2009.

\bibitem{Sotiriou:2009kx}
Thomas~P Sotiriou, Matt Visser, and Silke Weinfurtner.
\newblock {Quantum gravity without Lorentz invariance}.
\newblock arXiv:0905.2798, 2009.

\bibitem{Sotiriou:2009vn}
Thomas Sotiriou, Matt Visser, and Silke Weinfurtner.
\newblock {Phenomenologically viable Lorentz-violating quantum gravity}.
\newblock arXiv:0904.4464, 2009.

\bibitem{Volovik:2009av}
G.~E. Volovik.
\newblock $z=3$ {Lifshitz-Ho\v{r}ava model and Fermi-point scenario of emergent
  gravity}.
\newblock arXiv:0904.4113, 2009.

\bibitem{Charmousis:2009mz}
Christos Charmousis, Gustavo Niz, Antonio Padilla, and Paul~M. Saffin.
\newblock {Strong coupling in Ho\v{r}ava gravity}.
\newblock arXiv:0905.2579, 2009.

\bibitem{Lu:2009em}
H.~Lu, Jianwei Mei, and C.~N. Pope.
\newblock {Solutions to Ho\v{r}ava Gravity}.
\newblock arXiv:0904.1595, 2009.

\bibitem{Unruh:1981bi}
W.~G. Unruh.
\newblock Experimental black hole evaporation.
\newblock {\em Phys. Rev. Lett.}, 46:1351--1353, 1981.

\bibitem{Barcelo:2005ln}
Carlos Barcel\'o, Stefano Liberati, and Matt Visser.
\newblock Analogue gravity.
\newblock {\em Living Rev. Rel.}, 8:12, 2005.

\bibitem{Volovik:2003jn}
Grigory~E. Volovik.
\newblock {\em The universe in a {Helium} droplet}.
\newblock International Series of Monographs on Physics. Oxford University
  Press, Jun 1 2003.

\bibitem{Matt.-Visser:2002ot}
M.~Novello, M.~Visser, and G.~Volovik.
\newblock {\em {Artificial Black Holes}}.
\newblock World Scientific, Singapore; River Edge, U.S.A., 2002.

\bibitem{Schutzhold:2007aa}
Ralf Sch{\"{u}}tzhold and W.~G. Unruh.
\newblock Quantum analogues: From phase transitions to black holes and
  cosmology.
\newblock {\em Lecture Notes in Physics}, 718:300 p., 2007.

\bibitem{Weinfurtner:2007qy}
Silke Weinfurtner.
\newblock Emergent spacetimes.
\newblock arXiv:0711.4416, 2007.

\bibitem{Weinfurtner:2007aa}
Silke Weinfurtner, Angela White, and Matt Visser.
\newblock Trans-{Planckian} physics and signature change events in {Bose} gas
  hydrodynamics.
\newblock {\em Physical Review D}, 76:124008, 2007.

\bibitem{Weinfurtner:2006iv}
Silke Weinfurtner, Stefano Liberati, and Matt Visser.
\newblock Modelling {Planck}-scale {Lorentz} violation via analogue models.
\newblock {\em J. Phys. Conf. Ser.}, 33:373--385, 2006.

\bibitem{Liberati:2006kw}
Stefano Liberati, Matt Visser, and Silke Weinfurtner.
\newblock Analogue quantum gravity phenomenology from a two-component
  {Bose}--{Einstein} condensate.
\newblock {\em Class. Quant. Grav.}, 23:3129--3154, 2006.

\bibitem{Weinfurtner:2006nl}
Silke Weinfurtner, Stefano Liberati, and Matt Visser.
\newblock Analogue model for quantum gravity phenomenology.
\newblock {\em J. Phys.}, A39:6807--6814, 2006.

\bibitem{Liberati:2006sj}
Stefano Liberati, Matt Visser, and Silke Weinfurtner.
\newblock Naturalness in emergent spacetime.
\newblock {\em Phys. Rev. Lett.}, 96:151301, 2006.

\bibitem{White:2008xr}
Angela White, Silke Weinfurtner, and Matt Visser.
\newblock {Signature change events: A challenge for quantum gravity?}
\newblock arXiv:0812.3744, 2008.

\bibitem{Weinfurtner:2007ab}
Silke Weinfurtner, Piyush Jain, Matt Visser, and C.~W. Gardiner.
\newblock Cosmological particle production in rainbow spacetimes.
\newblock arXiv:0801.2673v1, 2007.

\bibitem{Pethick:2001aa}
C.~J. Pethick and H.~Smith.
\newblock {\em {Bose--Einstein} Condensation in Dilute Gases}.
\newblock Cambridge University Press, 2001.

\bibitem{Gardiner:2000aa}
Crispin Gardiner and Peter Zoller.
\newblock {\em Quantum Noise}.
\newblock Springer Series in Synergetics. Springer, Berlin, 2nd edition, 2000.

\bibitem{Barcelo:2002dp}
Carlos Barcel\'o, Stefano Liberati, and Matt Visser.
\newblock Refringence, field theory, and normal modes.
\newblock {\em Class. Quant. Grav.}, 19:2961--2982, 2002.

\bibitem{Visser:1993tk}
Matt Visser.
\newblock Acoustic propagation in fluids: {An} unexpected example of
  {Lorentzian} geometry.
\newblock {\em gr-qc/9311028}, 1993.

\bibitem{Visser:1997ux}
Matt Visser.
\newblock {Acoustic black holes: Horizons, ergospheres, and Hawking radiation}.
\newblock {\em Class. Quant. Grav.}, 15:1767--1791, 1998.

\bibitem{Inouye:1998aa}
S.~Inouye, M.~R. Andrews, J.~Stenger, H.~J. Miesner, D.~M. Stamper-Kurn, and
  W.~Ketterle.
\newblock Observation of {Feshbach} resonances in a {Bose}--{Einstein}
  condensate.
\newblock {\em Nature}, 392:152--154, 1998.

\bibitem{Donley:2001aa}
Elizabeth~A. Donley, Neil~R. Claussen, Simon~L. Cornish, Jacob~L. Roberts,
  Eric~A. Cornell, and Carl~E. Wieman.
\newblock Dynamics of collapsing and exploding {Bose}--{Einstein} condensates.
\newblock {\em Nature}, 412(6844):295--299, 2001.

\bibitem{Roberts:2001aa}
J.~L. Roberts, N.~R. Claussen, S.~L. Cornish, E.~A. Donley, E.~A. Cornell, and
  C.~E. Wieman.
\newblock Controlled collapse of a {Bose}--{Einstein} condensate.
\newblock {\em Phys. Rev. Lett.}, 86(19):4211--4214, May 2001.

\bibitem{Calzetta:2005yk}
E.~A. Calzetta and B.~L. Hu.
\newblock Early universe quantum processes in {BEC} collapse experiments.
\newblock {\em Int. J. Theor. Phys.}, 44:1691--1704, 2005.

\bibitem{Calzetta1:2003xb}
E.~A. Calzetta and B.~L. Hu.
\newblock Bose--{Einstein} condensate collapse and dynamical squeezing of
  vacuum fluctuations.
\newblock {\em Phys. Rev. A}, 68(043625), 2003.

\bibitem{Hu:2003}
B.~L. Hu and E.~A. Calzetta.
\newblock {BEC} collapse, particle production and squeezing of the vacuum.
\newblock arXiv:cond-mat/0208569, 2002.

\bibitem{Jain:2006ki}
Piyush Jain, Silke Weinfurtner, Matt Visser, and Crispin Gardiner.
\newblock {Analogue model of an expanding FRW universe in} {Bose}--{Einstein
  condensates}: {Application of the classical field method}.
\newblock {\em Physical Review A}, 76:033616, 2007.

\end{thebibliography}
\end{document}